\documentclass[aps,preprint,showpacs,showkeys,preprintnumbers,eqsecnum,amsmath,amssymb]{revtex4}

\usepackage{graphicx}
\usepackage{dcolumn}
\usepackage{mathrsfs}
\usepackage{bm}
\usepackage{color,xspace}
\usepackage[small,compact,raggedright]{titlesec}

\def\ra{\rangle}
\def\la{\langle}

\def\no{\nonumber}
\def\bea{\begin{eqnarray}}
\def\eea{\end{eqnarray}}
\def\be{\begin{equation}}
\def\ee{\end{equation}}

\begin{document}
\title{Entropic uncertainty relation in de Sitter space}

\author{Lijuan Jia, Zehua Tian and
Jiliang {Jing}\footnote{Corresponding author, Electronic address:
jljing@hunnu.edu.cn}}

\affiliation{Department of Physics, and Key Laboratory of Low Dimensional Quantum Structures and Quantum Control of Ministry of Education, Hunan Normal University, Changsha, Hunan 410081, P. R. China}
\begin{abstract}

The uncertainty principle restricts our ability to simultaneously predict the measurement outcomes of two incompatible observables of a quantum particle. However, this uncertainty could be reduced and quantified by a new Entropic Uncertainty Relation (EUR). By the open quantum system approach, we explore how the nature of de Sitter space affects the EUR. When the quantum memory $A$ freely falls in the de Sitter space, we demonstrate that the entropic uncertainty acquires an increase resulting from a thermal bath with the Gibbons-Hawking temperature. And for the static case, we find that the temperature coming from both the intrinsic thermal nature of the de Sitter space and the Unruh effect associated with the proper acceleration of $A$ also brings effect on entropic uncertainty, and the higher temperature, the greater uncertainty and the quicker the uncertainty reaches the maxima value. And finally the possible mechanism behind this phenomenon is also explored.

\end{abstract}


\maketitle

\section{Introduction}

Heisenberg's uncertainty relation \cite{H1}, which lies at the heart of understanding quantum mechanics, provides a dramatic illustration of a qualitative distinction between quantum and classical physics. This principle states that there is general irreducible lower bound on the uncertainty in the result of simultaneous measurement of two conjugate quantum mechanical variables, such as position and momentum, and more precisely, the product of the uncertainties in such two measurements is at least of order $\hbar$, or equivalently, there is an upper bound on the accuracy with which the values of noncommunity observables can be simultaneously prepared.

Due to the appearance of information theory, a more natural choice to measure uncertainty is based on entropy \cite{H2,H3,H4,H5,H6}. For non-commuting observables $Q$ and $R$, Deutsch \cite{H3} has described the relation as
\be\label{u1}
S_H(Q)+S_H(R)\geq -2 \log_2 \frac{1}{2} (1+c),
\ee
where $S_H=-\sum_jp(j)\log_2p(j)$, $Q$ and $R$ denote two Hermitian operators representing physical observables in an $N$-dimensional Hilbert space with $\{|a_j\rangle\}$ and $\{|b_j\rangle\}$ ($j$=1,...,$N$) the respective complete sets of normalized eigenvectors and $c=\mbox{max}_{i,j}|\la a_i| b_j\ra|$. Particularly, Kraus \cite{H4} suggested that this relation may be improved to
\be
S_H(Q)+S_H(R)\geq \log_2 \frac{1}{c^2}.
\label{eur1}
\ee
A distinct advantage of these relations, (\ref{u1}) and (\ref{eur1}), over the standard deviations is that the right-hand side is independent of the state of the system when the two measurements $Q$ and $R$ do not share any common eigenvector, i.e, it gives a fixed lower bound. So, they provide us a more general framework to quantify uncertainty.

However, using previously determined quantum information about the measured system, the above uncertainty bound could be violated. To overcome this defect, recently Refs. \cite{EUR1,EUR2} have given a stronger Entropic Uncertainty Relation (EUR) based on conditional entropy theoretically. Furthermore, several experiments \cite{EUR3,EUR4} have been performed to confirm this EUR.
For an entangled quantum system consisting of interesting particle $B$ and its quantum memory $A$, which is a device that might be available in the not-too-distant future and could store the information
of the entanglement between particles \cite{Mario Berta}, the conditional entropy EUR is shown as
\be
S_v(Q|A)+S_v(R|A)\geqslant\log_2\frac{1}{c^2}+S_v(B|A),
\label{eur2}
\ee
where $S_v(B|A)=S_v(\rho_{AB})-S_v(\rho_{A})$ is the conditional von Neumann entropy. In the extreme case, i.e., $A$ and $B$ are maximally entangled, it is able to predict the outcomes precisely. On the other hand, if $A$ and $B$ are not entangled, the bound in (\ref{eur1}) is recovered. The generalization of the EUR (\ref{eur2}) to R\'enyi entropy has also been given \cite{EUR5,EUR6}. Other studies from various views can be found in \cite{EUR7,EUR8,EUR9}.

It is well known that every quantum system, whatever it is, in a realistic regime is inevitably in contact with environments. As a result, the considered quantum system has to suffer a decoherence or dissipation. So
the nature of environment plays a key role in dominating the evolution of the quantum system, as well as the quantum-memory-assisted EUR \cite{APP1}. Besides the generally studied noisy channels, such as bit flip, noises resulting from the motions of observers or gravitational field are also a very important branch of quantum
noisy channels. Especially, such noises directly relate to the nature of spacetime, such as Hawing effect, and allow
us to incorporate the concepts of quantum information into relativistic settings. This combination has recently resulted in an entire novel field of physics, relativistic quantum information \cite{Mann}. Its aim is to answer questions about the overlap of relativity and the manipulation of information stored in quantum system, provide us a more completely frame to understand quantum information theoretically, and more importantly be a guidance for future realistic quantum information assignments in curved spacetime. So such works are very meaningful. J. Feng et al in recent work \cite{jun feng} has studied how the Unruh effect affects the EUR, which is the first try to discuss how the motion of the observer affects the Heisenberg's limit. However, their analysis is confined in the flat spacetime and the effects result only from the motion of observer. Therefore, it remains interesting to see what happens to the EUR if the quantum system is placed in a curved spacetime rather than a flat one.

In this paper, we will study the EUR (\ref{eur2}) under the decoherence rooting in vacuum field fluctuation in the de Sitter space. The reason for special attention to the de Sitter space in recent years is that our current observations, together with the theory of inflation, suggest that our universe may approach the de Sitter geometries in the far past and the far future. And a duality may exist between quantum gravity on the de Sitter space and a conformal field theory living on the boundary identified with the timelike infinity of it \cite{Strominger}. So, many fields, such as fields quantization \cite{dsp3,dsp4,dsp5,dsp6}, Lamb shift \cite{zhou1} and spontaneous excitation of atom \cite{Yu}, have been studied in this special curved spacetime, and it is necessary for us to focus on this spacetime to study EUR.

The model we will study is constructed like this: the quantum memory $A$, which interacts with quantized conformally coupled massless scalar fields in the de Sitter-invariant vacuum, freely falls or keeps static in spacetime. Another particle $B$, isolated from external field, denotes the system to be measured. They initially entangles with each other maximally. No matter which case, we find the quantum memory $A$, due to the fluctuation of vacuum field, will suffer from the thermal effect of spacetime, which acts as a type of noise channel. Under this noise channel the quantum information stored initially in $A$ would be decreased, thus, leading to an inevitably increase of the uncertainty on the outcome of measurements performed by observer. Along with the evolution of the quantum state, the uncertainty eventually achieve a finite maximal value. This phenomenon is essentially similar to that reported in \cite{APP1} where entanglement transfers between the quantum system and its environment.

Our paper is organized as follows: after briefly reviewing evolution of the quantum system and simply representing the entropic uncertainty relation in section \ref{section 2}, we calculate and discuss the entropic uncertainty with particle $A$ freely falling in the de Sitter space in section \ref{section 3} and with particle $A$ keeping static in section \ref{section 4}. Then, we will try to explain the possible mechanism in section V before summarizing our conclusions in section \ref{section 5}.

\section{Evolution of quantum system and entropic uncertainty}
\label{section 2}

We will discuss the evolution of quantum system and the entropic uncertainty.

\subsection{Evolution of quantum system}
Let's start with the Hamiltonian of the system containing particle and external field, which can be expressed as
\begin{eqnarray}\label{Hamiltonian}
H=H_s+H_\phi+H_I,
\end{eqnarray}
where $H_s$ and $H_\phi$ are the Hamiltonian of the particle and scalar field, respectively, and $H_I$ represents their interaction. For simplicity, we take a two-level particle with Hamiltonian $H_s=\frac{1}{2}\omega_0\sigma_z$, where $\omega_0$ is the energy level spacing, and $\sigma_z$ is the Pauli matrix. We suppose that the Hamiltonian representing the interaction between particle and scalar field is $H_I=\mu(\sigma_++\sigma_-)\phi(x(\tau))$, in which $\mu$ is a coupling constant, $\sigma_+$ ($\sigma_-$) is the rasing (lowering) operator, and $\phi(x)$ corresponds to the scalar field operator, which is conformally coupled to de Sitter space. Although there are kinds of coupling ways, only the conformal coupling preserves the de Sitter-invariant vacuum states, the others are those which break de Sitter vacuum invariance \cite{A}.

We assume that initially the two-level particle and field states are uncorrelated so that
\begin{eqnarray}
\rho_{tot}=\rho(0)\otimes|0\rangle\langle0|,
\end{eqnarray}
where $\rho(0)$ is the reduced density matrix of the two-level particle, and $|0\rangle$ is the vacuum state of the field. For the total system, its equation of motion in Schrodinger picture is
\begin{eqnarray}\label{motion equation}
\frac{\partial\rho_{tot}(\tau)}{\partial\tau}=-i[H,\rho_{tot}(\tau)],
\end{eqnarray}
where $\tau$ is the proper time of the two-level particle. In the limit of weak coupling, the evolution of the reduced density matrix $\rho(\tau)$, after some calculations, can be written in the Lindblad form \cite{Benatti,jason doukas}
\begin{eqnarray}\label{Lindblad equation}
\frac{\partial\rho(\tau)}{\partial\tau}=-i[H_{eff},\rho(\tau)]+\sum^3_{j=1}\left[2 L_j\rho
L_j^\dagger-\left\{L_j^\dagger L_j,\rho\right\}\right],
\end{eqnarray}
where $\{x,y\}=xy+yx$ denotes an anticommutator, $H_{eff}$ is the effective Hamiltonian, and $L_j$ are the Lindblad operators, which are given by
\begin{eqnarray}
\nonumber
H_{eff}&=&\frac{1}{2}\Omega\sigma_z=\frac{1}{2}\{\omega_0+\mu^2\mathrm{Im}(\Gamma_++\Gamma_-)\}\sigma_z
\\L_1 &=&\sqrt{\frac{\gamma_-}{2}} \sigma_-,\
L_2 =\sqrt{\frac{\gamma_+}{2}} \sigma_+,\
L_3 =\sqrt{\frac{\gamma_z}{2}}\sigma_z,
\end{eqnarray}
with
\begin{eqnarray}
\nonumber
\gamma_\pm&=&2\mu^2\mathrm{Re}\Gamma_\pm=\mu^2\int^{+\infty}_{-\infty}e^{\mp i\omega_0s}G^+(s-i \epsilon)ds,
\ \ \ \ \
\nonumber
\gamma_z=0,
\end{eqnarray}
here $G^+(x-x')=\langle0|\phi(x)\phi(x')|0\rangle$ and $s=\tau-\tau'$.

In our setup, we take two particles, keeping one of them particle $B$ isolated from external field while the other particle $A$ interacts with the environment. It is needed to note that this model is in structural similarity to a bipartite quantum system in quantum information theory, with one subsystem in interaction with external environment, and the other isolated from that. In this regard, let's note that this model has been used to discuss the loss of spin entanglement for accelerated electrons in electric magnetic fields \cite{jason doukas}, and the entanglement of two qubits in a relativistic orbit \cite{jason doukas1}. Since $\rho$ spans a sixteen dimensional vector space and the direct product of Pauli matrices including the identity, $\{\sigma_i\otimes\sigma_j|{i,j}\in{0,\cdots,3}\}$, form sixteen linearly independent vectors, we can expand the density matrix as
\begin{equation}\label{eqn:Arbintialdensity}
\rho_{AB}=\sum_{i=0}^3\sum_{j=0}^3r_{ij}\sigma_i\otimes\sigma_j.
\end{equation}
A nice property about this choice of basis is that the expansion coefficients $r_{ij}$ are real, which follows from the hermiticity of the Pauli matrices and density operator.
Substituting (\ref{eqn:Arbintialdensity}) into Eq. (\ref{Lindblad equation}) we have
\begin{align}
\frac{dr_{ij}}{d\tau}\sigma_{i}\otimes\sigma_{j}&=-\frac{i}{2}\Omega  r_{ij}(\sigma_z\sigma_i-\sigma_i\sigma_z)\otimes\sigma_j + \frac{\gamma_-}{2} r_{ij}\left[ 2
\sigma_-\sigma_{i}\sigma_+-\sigma_+\sigma_-\sigma_{i}-\sigma_{i}\sigma_+\sigma_-\right]\otimes \sigma_{j}\nonumber\\&+\frac{\gamma_+}{2}r_{ij}\left[ 2
\sigma_+\sigma_{i}\sigma_--\sigma_-\sigma_+\sigma_{i}-\sigma_{i}\sigma_-\sigma_+\right]\otimes\sigma_{j},
\end{align}
which after a little algebras gives sixteen first order linear differential equations
\bea
\dot{r}_{0j}(\tau)&=&0,\nonumber\\
\dot{r}_{1j}(\tau)&=&-\tfrac{1}{2}(\gamma_-+\gamma_+)r_{1j}(\tau)-\Omega r_{2j}(\tau),\nonumber\\
\dot{r}_{2j}(\tau)&=&-\tfrac{1}{2}(\gamma_-+\gamma_+)r_{2j}(\tau)+\Omega r_{1j}(\tau),\nonumber\\
\dot{r}_{3j}(\tau)&=&(\gamma_+-\gamma_-)r_{0j}(\tau)-(\gamma_-+\gamma_+)r_{3j}(\tau),\label{eqn:rdot1}
\eea
where dots imply differentiation with respect to $\tau$.
The solutions to these equations are found to be
\begin{align}\label{eqn:solution}
r_{0j}(\tau)&=r_{0j}(0),\nonumber \\
r_{1j}(\tau)&=r_{1j}(0)e^{-\tfrac{1}{2}a\tau}\cos{\Omega\tau}-r_{2j}(0)e^{-\tfrac{1}{2}a\tau}\sin{\Omega\tau},\nonumber \\
r_{2j}(\tau)&=r_{2j}(0)e^{-\tfrac{1}{2}a\tau}\cos{\Omega\tau}+r_{1j}(0)e^{-\tfrac{1}{2}a\tau}\sin{\Omega\tau}, \nonumber\\
r_{3j}(\tau)&=r_{3j}(0)e^{-a\tau}+\tfrac{b}{a}r_{0j}(0)(1-e^{-a\tau}),
\end{align}
where $a=\gamma_++\gamma_-,b=\gamma_+-\gamma_-$.

\subsection{Entropic uncertainty}

We consider the bipartite system, particle $A$ and particle $B$, initially share a maximally entangled Bell state
\begin{equation}\label{eqn:rhobell}
\rho_{AB}=\frac{1}{4}(\sigma_0\otimes\sigma_0+\sigma_1\otimes\sigma_1
-\sigma_2\otimes\sigma_2+\sigma_3\otimes\sigma_3).
\end{equation}
Then the corresponding time-dependent density matrix is
\begin{align}\label{eqn:evolution}
\rho_{AB}=\frac{1}{4}\{\sigma_0^A\otimes\sigma_0^B+e^{-\frac{1}{2} a \tau}\cos{\Omega\tau}\sigma_1^A\otimes\sigma_1^B-e^{-\frac{1}{2} a \tau}\cos{\Omega\tau}\sigma_2^A\otimes\sigma_2^B+e^{-a \tau}\sigma_3^A\otimes\sigma_3^B\nonumber\\
+e^{-\frac{1}{2} a \tau}\sin{\Omega\tau}\sigma_1^A\otimes\sigma_2^B+e^{-\frac{1}{2} a \tau}\sin{\Omega\tau}\sigma_2^A\otimes\sigma_1^B+\frac{b}{a}(1-e^{-a \tau})\sigma_3^A\otimes\sigma_0^B\}.
\end{align}

Now we assume a measurement is performed on the particle $B$ of (\ref{eqn:evolution}) in terms of one of the Pauli operators $\sigma_i$. The reason why we select the $\sigma_i$ to be measured is that they are the spin polarization components of the two-level atom, we can not simultaneously have complete information about both the observables $\sigma_1$ and $\sigma_3$, i.e, they have met the conditions that EUR (1.3) requires. Moreover, the Pauli operators are Hermitian, so $\sigma_1$ and $\sigma_3$ are two positive operator valued measurements (POVMs) acting on particle $B$.

After the measurement, the new post measurement states $\sum_x(\mathbf{1}\otimes\Pi_x)\rho_{AB}(\mathbf{1}\otimes\Pi_x)$, where $\Pi_x=|\psi_x\rangle\langle\psi_x|$ and ${|\psi_x\rangle}$ are the eigenstates of the observable $\sigma_i$, are given by
\begin{equation}
\rho_{A\sigma_1}=\frac{1}{4}\{\sigma_0^A\otimes\sigma_0^B+e^{-\frac{1}{2} a \tau}\cos{\Omega\tau}\sigma_1^A\otimes\sigma_1^B+e^{-\frac{1}{2} a \tau}\sin{\Omega\tau}\sigma_2^A\otimes\sigma_1^B+\frac{b}{a}(1-e^{-a \tau})\sigma_3^A\otimes\sigma_0^B\},
\end{equation}
\begin{equation}
\rho_{A\sigma_2}=\frac{1}{4}\{\sigma_0^A\otimes\sigma_0^B-e^{-\frac{1}{2} a \tau}\cos{\Omega\tau}\sigma_2^A\otimes\sigma_2^B+e^{-\frac{1}{2} a \tau}\sin{\Omega\tau}\sigma_1^A\otimes\sigma_2^B+\frac{b}{a}(1-e^{-a \tau})\sigma_3^A\otimes\sigma_0^B\},
\end{equation}
\begin{equation}
\rho_{A\sigma_3}=\frac{1}{4}\{\sigma_0^A\otimes\sigma_0^B+e^{-a \tau}\sigma_3^A\otimes\sigma_3^B+\frac{b}{a}(1-e^{-a \tau})\sigma_3^A\otimes\sigma_0^B\}.
\end{equation}
The eigenvalues can be easily calculated and the corresponding von Neumann entropy is
\bea
\no
S_v(\rho_{A\sigma_i})&=& 2 S_{H\mbox{\tiny bin}}(\lambda_i) \quad(i=1,2)\label{bent1},
\\
S_v(\rho_{A\sigma_3})&=& S_{H\mbox{\tiny bin}}(\eta_1)+S_{H\mbox{\tiny bin}}(\eta_2)\label{bent3},
\eea
with
\begin{eqnarray}
\no
\lambda_i&=&\frac{1}{4}\{1-[e^{-a \tau}+\frac{b^2}{a^2}(1-e^{-a \tau})^2]^{\frac{1}{2}}\}~~~(i=1,2),
\no \\
\eta_1&=&\frac{1}{4}[1+\frac{b}{a}+(1-\frac{b}{a})e^{-a \tau}],
\no \\
\eta_2&=&\frac{1}{4}(1+\frac{b}{a})(1-e^{-a \tau}),
\end{eqnarray}
where $S_{H\mbox{\tiny bin}}(p)=-p\log_2p-(\frac{1}{2}-p)\log_2(\frac{1}{2}-p)$.

Since $\rho_A=\mbox{Tr}_{\mbox{\tiny B}}\rho_{AB}$, the associated entropy is binary $S_v(\rho_A)=S'_{H\mbox{\tiny bin}}(\frac{1}{2}+\frac{b}{2a}(1-e^{-a \tau}))$, and $S'_{H\mbox{\tiny bin}}(p)=-p\log_2p-(1-p)\log_2(1-p)$ denoted as the binary entropy. For a particular measurement of $\sigma_1$ and $\sigma_3$ by observer, we can give the left-hand side (LHS) of (\ref{eur2}), represented as a uncertainty $U$ (one can gain similar result for the measurement on $\sigma_1$ and $\sigma_2$ ). $U(\sigma_1,\sigma_3)=S_v(\sigma_1|A)+S_v(\sigma_3|A)=S_v(\rho_{A\sigma_1})+S_v(\rho_{A\sigma_3})-2S_v(\rho_{A})$, where the sum of $S_v(\sigma_1|A)+S_v(\sigma_3|A)$ actually means the uncertainty about the outcomes of measurement $\sigma_1$ and $\sigma_3$ simultaneously given information stored in a quantum memory $A$. Hence, after a simple and straightfoward calculations, we can get the expression
\bea
U(\sigma_1,\sigma_3)=2 S_{H\mbox{\tiny bin}}(\lambda_1)+ S_{H\mbox{\tiny bin}}(\eta_1)+S_{H\mbox{\tiny bin}}(\eta_2)-2S'_{H\mbox{\tiny bin}}(\frac{1}{2}+\frac{b}{2a}(1-e^{-a \tau})).
\label{lhs1}
\eea

We now investigate the right-hand side (RHS) of (\ref{eur2}). Once the measurement choice has been determined, the complementarity $c$ of the observables $\sigma_j$ and $\sigma_k$ is always $1/\sqrt{2}$. The conditional entropy now is $S_v(B|A)=S_v(\rho_{AB})-S_v(\rho_A)$. The eigenvalues of (\ref{eqn:evolution}) are
\begin{eqnarray}
\xi_1&=&\frac{1}{4}(1-\frac{b}{a})(1-e^{-a \tau}),\no \\
\xi_2&=&\frac{1}{4}(1+\frac{b}{a})(1-e^{-a \tau}),\no \\
\xi_3&=&\frac{1}{4}(1+e^{-a \tau}-[4e^{-a \tau}+\frac{b^2}{a^2}(1-e^{-a \tau})^2]^{\frac{1}{2}}),\no \\
\xi_4&=&\frac{1}{4}(1+e^{-a \tau}+[4e^{-a \tau}+\frac{b^2}{a^2}(1-e^{-a \tau})^2]^{\frac{1}{2}}).\no
\end{eqnarray}
Denoting the RHS of (\ref{eur2}) as $U_b$, we have
\be
U_b=-\sum_{i=1}^4 \xi_i \log_2 \xi_i-S'_{H\mbox{\tiny bin}}(\frac{1}{2}+\frac{b}{2a}(1-e^{-a \tau}))+1.
\label{rhs1}
\ee

Expressions (\ref{lhs1}) and (\ref{rhs1}) are the ones that we actually need. For different evolutions, parameters $a$ and $b$ have different values, thus, we can obtain different entropic uncertainty. In the following, we will calculate the entropic uncertainty for two special cases, one is that quantum memory $A$ freely falls in the de Sitter space, the other corresponds to that $A$ keeps static in the de Sitter space.

\section{EUR with quantum memory $A$ freely falling in de Sitter space}
\label{section 3}

We now consider particle $A$ freely falls and interacts with a quantized conformally coupled massless scalar field in the de Sitter space. There are several different coordinate systems can be chosen to characterize the de Sitter space \cite{dsp2,dsp15}. Here we choose to work with the global coordinate system $(t,\chi,\theta,\phi)$ under which the freely falling particle $A$ is comoving with the expansion, and the corresponding line element is
\begin{eqnarray}\label{line element 1}
ds^2=dt^2-\alpha^2\cosh^2(t/\alpha)[d\chi^2+\sin\chi^2 (d\theta^2+\sin\theta^2d\phi^2)]
\end{eqnarray}
with $\alpha=\sqrt{\frac{3}{\Lambda}}$, where $\Lambda$ is the cosmological constant. The parameter $t$ is often called the world or cosmic time. The scalar curvature of the spacetime is $R=12\alpha^{-2}$. If $-\infty<t<\infty$, $0\leq\chi\leq\pi$, $0\leq\theta\leq\pi$, $0\leq\phi\leq2\pi$, the coordinate covers the whole de Sitter manifold \cite{dsp2,dsp3,dsp15}.

The canonical quantization of scalar
field with this metric has been done in \cite{dsp3,dsp4,dsp5,dsp6}, in the massless and conformal coupling limit, the Wightman function
for a freely-falling particle $A$ can be simplified to be
\begin{eqnarray}\label{Wightman function 1}
G^+(x-x')=-\frac{1}{16\pi^2\alpha^2\sinh^2(\frac{\tau-\tau'}{2\alpha}-i\epsilon)},
\end{eqnarray}
where $\tau=t$. Therefore, we find
\begin{eqnarray}
a&=&\frac{\mu^2\omega_0}{2 \pi}\left(\frac{e^{2\pi\alpha\omega_0}+1} {e^{2\pi\alpha\omega_0}-1}\right),
\nonumber \\ b&=&-\frac{\mu^2\omega_0}{2 \pi},\nonumber\\
H_{eff}&=&
\frac{1}{2}\big\{\omega_0+\frac{\mu^2}{4\pi^2}\int^\infty_0d\omega
P(\frac{\omega}{\omega+\omega_0}-\frac{\omega}{\omega-\omega_0})
(1+\frac{2}{e^{2\pi\alpha\omega}-1})\big\}
\sigma_z, \label{Lam shift}
\end{eqnarray}
where the last term of $H_{eff}$ represents the Lamb shift \cite{zhou1} in the de Sitter space. Then, we can get $U(\sigma_1,\sigma_3)$ and $U_b$ for this case.

From Eq. (\ref{Lam shift}) we know that the freely falling particle $A$ in the de Sitter space feels a Gibbons-Hawking temperature $T_f=1/2\pi\alpha$, which clearly suggests that the intrinsic thermal nature of the de Sitter space exists. And we can find the uncertainty is related to temperature that particle $A$ feels. When $\alpha\rightarrow\infty$, i.e, $T_f=0$, it corresponds to particle $A$ freeing from the thermal effect. So what influence the temperature has on the entropic uncertainty, depicted in Fig. 1.

\begin{figure}[hbtp]
\includegraphics[width=.43\textwidth]{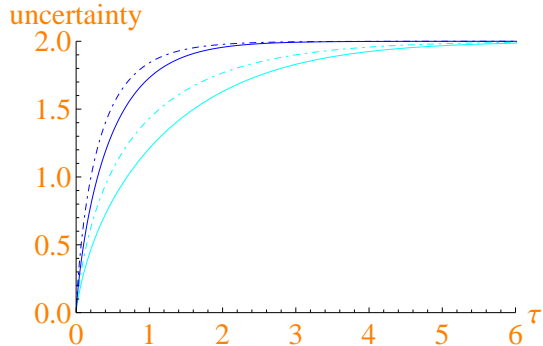}
\caption{the $U(\sigma_1,\sigma_3)$ (dashed line) and $U_b$ (solid line) as a function of proper time, $\tau$, in units of $\gamma_0^{-1}$, $\gamma_0=\frac{\mu^2\omega_0}{2\pi}$ is the spontaneous emission rate. The above two lines (blue colour) and the below two lines (cyan colour) correspond to $T_f=0.8$ (which is assumed) and $T_f=0$, in units of $\omega_0$, respectively.}
\label{EUR-1}
\end{figure}

As depicted in Fig. 1, for $\tau=0$, namely, $A$ and $B$ remain being maximally entangled, $U=U_b=0$, satisfying the EUR (\ref{eur2}), which means that one can predict the outcomes precisely. As time goes on, the uncertainty bound $U_b$ is lifted, meanwhile the uncertainty $U$ is also changed, but they still meet the EUR (\ref{eur2}). Finally, the uncertainty reaches a finite maxima value, about $2$, and $U= U_b$. Moreover, we observe that with $T_f=0.8$ the measurement outcome is more uncertain than that frees from thermal effect during the whole evolution, so we can arrive at the conclusion that the thermal nature of the de Sitter space surely increases the uncertainty. The possible mechanism behind this phenomenon will be studied in the following section.

\section{EUR with quantum memory $A$ keeping static in de Sitter space}
\label{section 4}

Next we will discuss under the same spacetime background for static particle $A$. On this occasion, we choose to work in the static coordinate system, and the line element is
\begin{eqnarray}\label{static coordinate}
ds^2=\big(1-\frac{r^2}{\alpha^2}\big)d\widetilde{t}^2-\big(1-\frac{r^2}{\alpha^2}\big)^{-1}dr^2
-r^2(d\theta^2+\sin\theta^2d\phi^2).
\end{eqnarray}
This metric possesses a event horizon at $r=\alpha$, generally named as cosmological horizon.
Note that the coordinates $(\widetilde{t},r,\theta,\phi)$ only cover part of the de Sitter space, just like the Rindler wedge in a flat spacetime. And the relation between the static and global coordinates system is
\begin{eqnarray}\label{static and global relation}
r=\alpha\cosh(t/\alpha)\sin\chi,~~~\tanh(\widetilde{t}/\alpha)=\tanh(t/\alpha)\sec\chi.
\end{eqnarray}
Obviously, the worldline $r=0$ in the static coordinate coincides with the worldline $\chi=0$ in the global coordinate, and an particle at rest with $r\neq0$ in the static coordinate will be accelerated relative to the observer at rest in the global coordinate with $\chi=0$, which is described by
\begin{eqnarray}\label{proper acceleraion}
a=\frac{r}{\alpha^2}\big(1-\frac{r^2}{\alpha^2}\big)^{-1/2}.
\end{eqnarray}

Similarly, in the static coordinates system by solving the field equation, one can find a set of complete eigenmodes \cite{dsp6,dsp7,dsp8,dsp9}. Defining a de Sitter-invariant vacuum, we can calculate the Wightman function for the massless conformally coupled scalar field, for static $A$, it is represented as \cite{dsp10,dsp11}
\begin{eqnarray}\label{wightman function 2}
G^+(x-x')=-\frac{1}{16\pi^2\kappa^2\sinh^2(\frac{\tau-\tau'}{2\kappa}-i\epsilon)},
\end{eqnarray}
where $\kappa=\sqrt{g_{00}}\alpha$ and $\tau=\sqrt{g_{00}}\widetilde{t}$. Then, we can easily acquire
\begin{eqnarray}
a&=&\frac{\mu^2\omega_0}{2 \pi}\left(\frac{e^{2\pi\kappa\omega_0}+1} {e^{2\pi\kappa\omega_0}-1}\right),
\nonumber \\ b&=&-\frac{\mu^2\omega_0}{2 \pi},\nonumber\\
H_{eff}&=&\frac{1}{2}\{\omega_0+\mu^2\mathrm{Im}(\Gamma_++\Gamma_-)\}\sigma_z
\nonumber \\
&=&\frac{1}{2}\big\{\omega_0+\frac{\mu^2}{4\pi^2}\int^\infty_0d\omega
P(\frac{\omega}{\omega+\omega_0}-\frac{\omega}{\omega-\omega_0})(1+\frac{2}{e^{2\pi\kappa\omega}-1})\big\}
\sigma_z. \label{Lam shift1}
\end{eqnarray}
In this case, static $A$ feels a temperature $T_s=1/2\pi\kappa$ in the de Sitter space. However it is needed to note that there remain differences from what was obtained in the case of the freely failing $A$ $(T_f=1/2\pi\alpha)$, we can connect
these two temperatures by
\begin{eqnarray}\label{relation between two temperature}
T^2_s=\big(\frac{1}{2\pi\alpha}\big)^2+\big(\frac{a}{2\pi}\big)^2=T^2_f+T^2_U,
\end{eqnarray}
in which the first term is the square of the Gibbons-Hawking temperature of the de Sitter space, and the second term is related to the Unruh temperature, which depends on proper acceleration described by Eq.(\ref{proper acceleraion}). Furthermore, $T_s$ varies with the rest position $r$, so located at different positions, particle $A$ feels a different temperature $T_s$, and the Von neumann entropy will have a remarkable change. Therefore, besides the proper time $\tau$, the entropic uncertainty also changes with respect to temperature $T_s$.

\begin{figure}[hbtp]
\includegraphics[width=.43\textwidth]{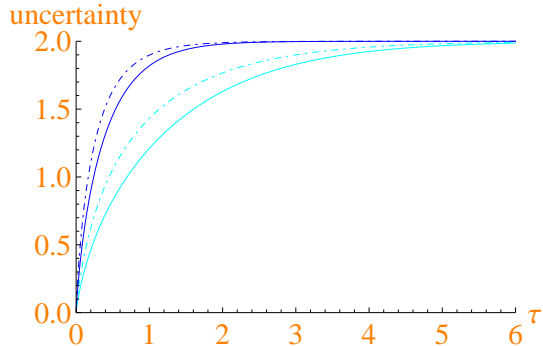}
\caption{the $U(\sigma_1,\sigma_3)$ (dashed line) and $U_b$ (solid line) as a function of proper time, $\tau$ (in units of $\gamma_0^{-1}$). The above two lines (blue colour) and the following two lines (cyan colour) correspond to $T_s=1.0$ (which is assumed) and $T_s=0$ ($\alpha\rightarrow\infty$ and $r=0$), in units of $\omega_0$, respectively.}
\label{EUR-2}
\end{figure}

The uncertainty is plotted in Fig. 2 as a function of the proper time. From the figure we find that the whole process is similar to the first case and meets the EUR (\ref{eur2}).
Furthermore, the uncertainty is greater under $T_s=1.0$ than that $T_s=0$ during the whole evolution. So we can arrive at the conclusion that the thermal nature of the de Sitter space surely increases the uncertainty. And we find that the higher temperature $T_s$, the quicker the uncertainty reaches the maxima value.

To illustrate how entropic uncertainty does vary with temperature $T_s$ more clearly, we select two points at $\tau=0.5$ and $\tau=1.2$ and depict the corresponding entropy changing with temperature $T_s$ in Fig. 3. From which we find that along with the increase of temperature, the uncertainty $U$ gradually increase and the uncertainty bound $U_b$ is also lifted. But they still satisfy the EUR (\ref{eur2}).

\begin{figure}[hbtp]
\includegraphics[width=.43\textwidth]{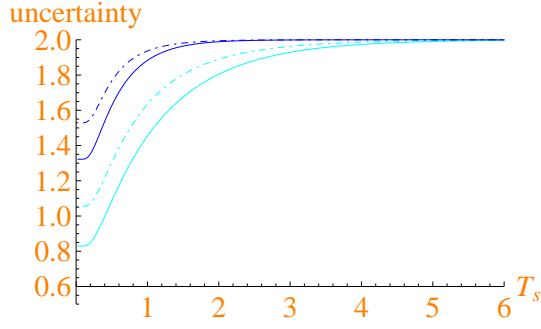}
\caption{
The $U(\sigma_1,\sigma_3)$ (dashed line) and $U_b$ (solid line) as a function of temperature $T_s$ (in units of $\omega_0$), the above two lines(blue colour) and the following two lines(cyan colour) correspond to $\tau=1.2$ and $\tau=0.5$ (in units of $\gamma_0^{-1}$), respectively.}
\label{EUR-3}
\end{figure}

\section{Relation between uncertainty and quantum correlation}

In this section, we will try to explain the phenomenon represented above from the perspective of quantum correlation. Although the proper time possessed by the freely falling particle $A$ distincts from which the static particle $A$ possesses, and so does the temperature. Conveniently, we mark the proper time by $\tau$ and temperature by $T$. We relate the lower bound of Eq. (\ref{eur2}) to the definition of discord: $D=-S_v(A|B)+min_{B_k}\Sigma_kq_kS_v(\rho_A^k)$, where $\rho_A^k=tr_B\{B_k\rho_{AB}B_k^\dagger\}/q_k$ is the resulting state after the complete measurement $\{B_k\}$ on particle $B$ and $q_k=tr_{AB}\{B_k\rho_{AB}B_k^\dagger\}$.

\begin{figure}[hbtp]
\includegraphics[width=.43\textwidth]{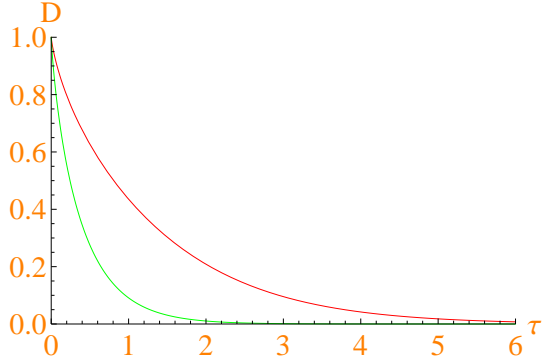}
\caption{the quantum correlation is plotted as a function of $\tau$ (in unites of $\gamma_0^{-1}$), the above line (red colour) and the following line (green colour) correspond to $T=0$ and $T=1.0$ (in units of $\omega_0$) respectively.}
\label{EUR-4}
\end{figure}

After some calculations, we depict the quantum correlation $D$ in Fig. 4. For $\tau=0$, the quantum correlation has the maxima value, corresponding to $U=U_b=0$ and particle $A$ and $B$ are maximally entangled that one can predict the outcomes precisely. As time goes on, it is likely that the decrease of quantum correlation makes the outcomes of two incompatible observables more uncertain and the lower bound $U_b$ lifted from Fig. 2 and Fig. 4. When the quantum correlation eventually vanishes, the uncertainty arrives at a maxima value, about 2. Furthermore, we find the thermal nature of the de Sitter space affects the value of $D$. Comparing quantum correlation under $T=1.0$ with that under $T=0$, the quantum correlation influenced by temperature $T=1.0$ decreases more for a period of time interval, which means that the thermal nature of the de Sitter space would surely increase the uncertainty.

To explain how the quantum correlation $D$ changes with respect to temperature $T$ more clearly, we depict it in Fig. 5. We find that the increase of temperature makes the quantum correlation decreased and the uncertainty increased.

Actually, QM uncertainty relations are most important precisely in those instances when they are saturated, so, next we will talk about the probability densities that saturate EUR. Seen from the figures 1 and 2, there are two cases in which the uncertainty relations are saturated, the maximally entangled case ($\tau$=0) and the totally decoherent case ($\tau$$\rightarrow$$\infty$). For $\tau$=0, the probability density is $\rho_{AB}=\frac{1}{4}(\sigma_0\otimes\sigma_0+\sigma_1\otimes\sigma_1-\sigma_2\otimes\sigma_2+\sigma_3\otimes\sigma_3)$. As $\tau$ is infinite, the probability density is $\rho_{AB}=\frac{1}{4}(\sigma_0^A\otimes\sigma_0^B+\frac{b}{a}\sigma_3^A\otimes\sigma_0^B)$. In addition, we can find out from the figure 3 that the uncertainty relations are saturated as the temperature $T_s$$\rightarrow$$\infty$, however, in order to meet the infinite temperature, from $T_s=\frac{1}{2\pi\kappa}=\frac{1}{2\pi\sqrt{\alpha^2-r^2}}$, if and only if $r=\alpha$, namely, particle $A$ stays at the event horizon.

\begin{figure}[hbtp]
\includegraphics[width=.43\textwidth]{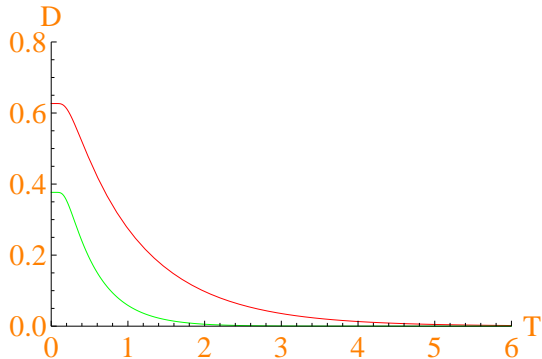}
\caption{the quantum correlation is plotted as a function of $T$ (in unites of $\omega_0$), the above line (red colour) and the following line (green colour) correspond to $\tau=0.5$ and $\tau=1.2$ (in units of $\gamma_0^{-1}$).}
\label{EUR-5}
\end{figure}

\section{Conclusions}
\label{section 5}

In the framework of open quantum system, we have explored how the nature of the de Sitter space affects the EUR. For a bipartite system, the quantum memory $A$ interacts with quantized conformally coupled massless scalar fields in the de Sitter-invariant vacuum, and the particle $B$ to be measured is initially entangled with $A$ and isolated from external environment. We have studied the evolution equation of quantum system and the entropic uncertainty for both freely falling and keeping static particle $A$. For the freely falling case, the quantum memory particle $A$ is immersed in a thermal bath with the Gibbons-Hawking temperature $T_f=1/2\pi\alpha$, which suggests that the intrinsic thermal nature of the de Sitter space exists. We find that the thermal nature of the de Sitter space could surely increase the uncertainty, and finally the uncertainty reaches a maxima value. For the static quantum memory particle $A$ in the de Sitter space, it feels a composite effect which contains the Gibbons-Hawking effect of the de Sitter space and the Unruh effect associated with the proper acceleration. The temperature that $A$ feels is a square root of the sum of the squared Gibbons-Hawking temperature and the squared Unruh temperature associated with proper acceleration. We also find that the thermal nature of the de Sitter space increases the uncertainty, and finally the uncertainty reaches a maxima value. Moreover, the uncertainty changes with respect to temperature $T_s$, the higher the temperature $T_s$ is, the greater the uncertainty is, and the quicker the uncertainty reaches the maxima value. For any cases, we find that all the processes meet the EUR (\ref{eur2}).

Finally, from the perspective of quantum correlation $D$, we have tried to explain the possible mechanism behind this phenomenon. We find that the decrease of quantum correlation may make the outcomes of two incompatible observables more uncertain and the lower bound $U_b$ lifted. When the quantum correlation eventually vanishes, the uncertainty arrives at a maxima value, about 2. And the increase of temperature makes the quantum correlation smaller, and the uncertainty becomes greater. In addition, we have talked about the probability densities that saturate EUR.

\section*{Acknowledgement}

This work was supported by the  National Natural Science Foundation
of China under Grant No. 11175065 and 11475061; the National Basic
Research of China under Grant No. 2010CB833004; the SRFDP under
Grant No. 20114306110003; Hunan Provincial Innovation Foundation For Postgraduate under Grant No CX2012B202.

\end{document}